\documentclass[12pt,a4paper]{article}

\usepackage[english]{babel}
\usepackage{amsmath,amsthm,amsfonts,bbm,graphicx,psfrag,cite}

\newcommand{\Xf}{\mathfrak{X}}
\newcommand{\Yf}{\mathfrak{Y}}
\renewcommand{\Re}{\mathrm{Re}}

\newcommand{\Tr}{\mathrm{Tr}}
\newcommand{\Id}{\mathbbm{1}}
\newcommand{\Or}{\mathcal{O}}
\newcommand{\Z}{\mathbbm{Z}}
\newcommand{\N}{\mathbbm{N}}
\newcommand{\R}{\mathbbm{R}}
\newcommand{\Pb}{\mathbbm{P}}
\newcommand{\dx}{\mathrm{d}}
\newcommand{\sgn}{\mathrm{sgn}}
\newcommand{\Ai}{\mathrm{Ai}}
\newcommand{\e}{\varepsilon}
\newcommand{\Dtn}[3]{\frac{\dx^{#3} #1}{\dx #2^{#3}}}
\newcommand{\Dt}[2]{\frac{\dx #1}{\dx #2}}
\newcommand{\Af}{{\cal A}_{\rm 1}}
\newcommand{\Ac}{{\cal A}_{\rm 2}}
\newcommand{\I}{{\rm i}}

\numberwithin{equation}{section}

\newtheorem{prop}{Proposition}[section]
\newtheorem{thm}[prop]{Theorem}
\newtheorem{lem}[prop]{Lemma}

\newtheorem{conj}{Conjecture}

\newenvironment{proofOF}[2]{\removelastskip\vspace{6pt}\noindent {\it Proof of #1.}~\rm#2}{\qed \par\vspace{6pt}}

\numberwithin{equation}{section}

\title{Fluctuation properties of the TASEP with periodic initial configuration}
\author{Alexei Borodin\thanks{California Institute of Technology, e-mail: borodin@caltech.edu},
Patrik L. Ferrari\thanks{Technische Universit\"at M\"unchen, e-mail: ferrari@ma.tum.de},\\
Michael Pr\"ahofer\thanks{Technische Universit\"at M\"unchen, e-mail: praehofer@ma.tum.de},
Tomohiro Sasamoto\thanks{Chiba University, e-mail: sasamoto@math.s.chiba-u.ac.jp}}

\date{12th December 2006}

\begin{document}
\maketitle \sloppy

\begin{abstract}
We consider the joint distributions of particle positions for the
continuous time totally asymmetric simple exclusion process
(TASEP). They are expressed as Fredholm determinants with a kernel
defining a signed determinantal point process. We then consider
certain periodic initial conditions and determine the kernel in
the scaling limit. This result has been announced first in a
letter by one of us~\cite{Sas05} and here we provide a
self-contained derivation. Connections to last passage directed
percolation and random matrices are also briefly discussed.
\end{abstract}

\section{Introduction}\label{SectIntro}
\emph{Continuous time TASEP.}  The totally asymmetric simple
exclusion process (TASEP) is one of the simplest interacting
stochastic particle systems. Its particles are on the lattice of
integers, $\Z$, with at most one particle at each site (exclusion
principle). The dynamics of the TASEP is defined as follows.
Particles jump on the neighboring right site with rate $1$
provided that the site is empty. This means that jumps are
independent of each other and take place after an exponential
waiting time with mean $1$, which is counted from the time instant
when the right neighbor site is empty.

On the macroscopic level the particle density, $u(x,t)$, evolves
deterministically according to the Burgers equation $\partial_t u
+ \partial_x(u(1-u))=0$~\cite{Rez91}. Therefore a natural question
is to focus on fluctuation properties and on large deviations,
which exhibit rather unexpected features. The fluctuations of the
integrated particle current and in the positions of particles are two
faces of the same coin as to be discussed later.

The fluctuations of particle positions are sensitive to the initial conditions. For example, one can consider particles initially positioned every second site, i.e., on $2\Z$. Another possibility would be to consider the stationary measure of the same density as initial condition, which is Bernoulli with density $1/2$. The scaling exponents for particle positions fluctuations are the same for the two initial conditions. However, the limiting distribution differ: as we will see, in the first case it is the GOE Tracy-Widom of random matrices~\cite{TW96} which differs from the stationary case~\cite{FS05a}. Thus it is of interest to understand which class of initial conditions leads to the same limit distribution.

The first result in this direction has been obtained for the step
initial condition. To be precise, denote by $x_k(t)$ the position
of particle $k$ at time $t$, where the $k$'s are integers
labelling the particles from right to left. The step initial
condition is then $x_k(0)=-k$, $k\in\N$. It has been studied by
Johansson~\cite{Jo00b} by means of a growth model. In terms of the
TASEP, the quantity analyzed is the large time asymptotic
fluctuations of the position of a given particle. For any fixed
$\alpha\in(0,1)$, the fluctuation of $x_{[\alpha t]}(t)$ are
asymptotically governed by the GUE Tracy-Widom distribution,
$F_2$, namely there are some $v=v(\alpha)$ and $b=b(\alpha)$ such
that
\begin{equation}\label{eq:joh1}
\lim_{t\to\infty}\Pb(x_{[\alpha t]}(t)\leq v(\alpha) t+s \, b(\alpha) t^{1/3})=F_2(s).
\end{equation}
The distribution $F_2$ first appeared in the context of the Gaussian Unitary Ensemble (GUE) of random matrices as the distribution of the largest eigenvalue in the limit of large matrix dimension~\cite{TW94}.

A natural question is to ask how the positions of different
particles are correlated, i.e., one considers for fixed but large
time $t$ the process $k\mapsto x_{k}(t)$. To illustrate known
results we focus at $k\sim t/4$, but the same holds with different
numerical coefficients for $k\sim \alpha t$,
$\alpha\in(0,1)$~\footnote{We choose $\alpha=1/4$ because then the
term linear in $t$ disappears in (\ref{eq1.2}), (\ref{eq:sas1}),
and (\ref{eqRescaling}).}. (\ref{eq:joh1}) tells us that the
fluctuations live on a $t^{1/3}$ scale and it turns out that the
position of two particles are, on the $t^{1/3}$ scale,
non-trivially correlated over a distance of order $t^{2/3}$. The
exponents $1/3$ and $2/3$ are indeed the ones of the KPZ
universality class~\cite{KPZ86}, to which the TASEP belongs.
Indeed, Johansson~\cite{Jo03b} proves a functional limit theorem in a discrete-time setting.
Its continuous-time version writes
\begin{equation}\label{eq1.2}
\lim_{t\to\infty}\frac{x_{[t/4+u(t/2)^{2/3}]}(t)-(-2u(t/2)^{2/3}+u^2(t/2)^{1/3})}{-(t/2)^{1/3}}=\Ac(u)
\end{equation}
where $\Ac$ is known as the Airy process, first discovered in the PNG model under droplet growth~\cite{PS02}.

Besides the step initial condition explained above, at least two
other situations are of particular interest. One is the
\emph{stationary} initial condition, for which the two-point
function of the TASEP is analyzed in~\cite{FS05a}. The second one
has \emph{deterministic} initial conditions leading to a
macroscopically uniform density, thus called \emph{flat} initial
conditions. The simplest realization is to set $x_k(0)=-2k$,
$k\in\Z$.

In~\cite{Sas05} an important new result has been obtained, making possible the asymptotic analysis for such initial conditions. First of all, as expected by universality, the position of a particle has fluctuations governed by the GOE Tracy-Widom distribution, $F_1$~\cite{TW96}. This result is a combination of~\cite{Sas05,FS05b} and states
\begin{equation}\label{eq:sas1}
\lim_{t\to\infty}\Pb(x_{[t/4]}(t)\leq - s t^{1/3})=F_1(2s).
\end{equation}
More importantly, in~\cite{Sas05} the analogue of the Airy process $\Ac$ for flat initial conditions, which we denote by $\Af$, is identified. It is a marginal of the signed determinantal point process with the extended kernel (\ref{eqKernelCompact}). Here signed refers to the non-positiveness of the measure (it does not define a probability measure). Explicitly, let $B_0(x,y)=\Ai(x+y)$ and let $\Delta$ be the one-dimensional Laplacian; then
\begin{equation}\label{eqKernelCompact}
K_{\rm F_1}(u_1,s_1;u_2,s_2)=-(e^{(u_2-u_1)\Delta})(s_1,s_2) \Id(u_2>u_1)+(e^{-u_1\Delta} B_0 e^{u_2\Delta})(s_1,s_2),
\end{equation}
or, equivalently as shown in Appendix~\ref{AppCompactKernel},
\begin{eqnarray}\label{eqKernelExpanded}
& &\hspace{-2em} K_{\rm F_1}(u_1,s_1;u_2,s_2)=-\frac{1}{\sqrt{4\pi (u_2-u_1)}}\exp\left(-\frac{(s_2-s_1)^2}{4 (u_2-u_1)}\right) \Id(u_2>u_1) \nonumber \\
& &\hspace{-2em} + \Ai(s_1+s_2+(u_2-u_1)^2) \exp\left((u_2-u_1)(s_1+s_2)+\frac23(u_2-u_1)^3\right).
\end{eqnarray}
This is in complete analogy with the Airy process $\Ac$, which is a marginal of the determinantal point process defined by the extended Airy kernel.

In Theorem~\ref{ThmJointCorr} we provide a derivation of the fact
that the joint distribution of particle positions is given by the
Fredholm determinant of a kernel. This is a general result which is then applied to the flat initial conditions $x_k(0)=-2k$, $k\in\Z$, see Theorem~\ref{ThmKernel}. The proper rescaling of
particle positions is
\begin{equation}\label{eqRescaling}
x_t^{\rm resc}(u)=-t^{-1/3}\big(x_{[t/4+ut^{2/3}]}(t)+2ut^{2/3}\big).
\end{equation}
With this rescaling, in the limit of large time $t$, the kernel
converges to $K_{\rm F_1}$ as shown in Theorem~\ref{ThmConvPt}.
There we show pointwise convergence. In a forthcoming
paper~\cite{BFP06} we will analyze a discrete time version of the
TASEP and strengthen our result to the convergence of the Fredholm
determinants. Such a stronger convergence would imply the
convergence of (\ref{eqRescaling}) to $\Af$ in the sense of
finite-dimensional distributions.

As a remark we want to point out that, while periodic initial condition does not seem to be accessible by previously known techniques, with the new construction both step and periodic initial conditions can be analyzed. The technique used so far is the reduction of the model to a determinantal point process via the Robinson-Schensted-Knuth correspondence. For further references and details on determinantal point processes we refer the reader to surveys~\cite{Lyo03,BKPV05,Sos06,Jo05} and the lecture notes~\cite{Spo05}.

\vspace{0.5em}\emph{Reformulation of the result.} The TASEP integrated current at position $x$ and time $t$, $J(x,t)$, is the number of particles which jumped from site $x$ to site $x+1$ during the time interval $[0,t]$. Let us label by $1$ the right-most particle starting at position $x_1(0)\leq x$. Then $\Pb(J(x,t)\geq s)=\Pb(x_s(t)\geq x+1)$. Thus the result of this paper translates directly to the integrated currents.

The TASEP can be mapped to last passage percolation on $\Z^2$ with
i.i.d.\ exponentially distributed random variables $\omega(i,j)$,
$i,j\in\Z$. $\omega(i,j)$ is the waiting time of particle number
$j$ to jump from position $i-j$ to $i-j+1$. There is a slight
switch in the point of view. For the TASEP one considers the
particle positions at fixed time $t$, while in last passage
percolation, one studies the last passage time from the origin to
points of a given lattice domain $\{i,j\in\Z^2,i+j=t\}$.
These two points of view are closely connected. They can be regarded
as taking different cross sections in the Bernoulli cone~\cite{Pra03}.
The problem considered in~\cite{Jo00b} is the point-to-point last passage percolation,
which corresponds to the step initial condition for the TASEP.
Flat initial conditions correspond to the point-to-line
percolation.

Finally, the same last passage percolation model can be seen as a
one-dimensional growth model~\cite{R81,Jo00b}, called discrete
polynuclear growth model, which serves as a discretized model for
KPZ growth~\cite{KPZ86}. KPZ growth is discussed in the books~\cite{BS95,Mea98}
but for a recent exposition of KPZ universality see~\cite{Pra03}.

\vspace{0.5em}\emph{Universality issue.} The TASEP also has
discrete time versions. One of these is the parallel update rule (see the review~\cite{Sch00}) and it is given as follows. At each time step particles jump to the
neighboring right site with probability $p\in(0,1)$, provided the
target site is empty. The jumps occurs independently and
simultaneously. There are two interesting limits of the discrete-time TASEP, namely $p\to 0$ and $p\to 1$.
The continuous-time TASEP is obtained by setting the time-unit to
$p$ and then take \mbox{$p\to 0$}. The limit $p\to 1$ yields the
so-called polynuclear growth (PNG) model, see e.g.~Section 2.1.5
of~\cite{FerPhD}. There, one has a height function $h$ on a
one-dimensional substrate, and flat initial condition for the
TASEP translates to growth starting with \mbox{$h(x,t=0)=0$}, also called
flat PNG.

By universality the process $\Af$ is expected to appear in the discrete-time TASEP and the PNG model as well. Universality has been confirmed for step initial conditions and the corresponding PNG model with droplet growth~\cite{Jo03b,PS02}. Moreover, for flat initial conditions, the limit process should be independent on the initial particle spacing. Results in a discrete-time version of the TASEP with different initial spacing will be presented in~\cite{BFP06}.

For the flat PNG model, it is known~\cite{BR00,PS00} that the height at one fixed point is GOE Tracy-Widom distributed in the limit of large time $t$. Thus, on the basis of the result for the TASEP with alternating initial conditions, see Theorem~\ref{ThmConvPt}, we can conjecture the behavior of the flat PNG model.
\begin{conj}\label{Cong1}
The properly rescaled height function of the PNG model with flat initial conditions converges, in the large time limit, to the process $\Af$.
\end{conj}
The scaling exponents are the same and the coefficients can be determined by matching with the PNG droplet case.

Finally, let us discuss the connection to random matrices. For the
TASEP with the step initial condition, the one-point asymptotic
distribution is the GUE Tracy-Widom distribution, $F_2$, and the
whole limit process is the Airy process $\Ac$.
The derivation uses an extension of the model
to a multi-layer version. The Airy process arises also for a GUE
matrix diffusion, the so-called $\beta=2$ Dyson's Brownian
Motion~\cite{Dys62}. The motion of the properly rescaled largest
eigenvalue converges to the Airy process. The connection extends
to finitely many of the largest eigenvalues which have the same
limiting behavior as the first top layers in the multi-layer PNG
model.

For the TASEP with flat initial condition, the one-point distribution
is the GOE Tracy-Widom distribution and the limit process is
$\Af$. At this point it is tempting to conjecture that the
evolution of the largest eigenvalue of a matrix which follows
$\beta=1$ Dyson's Brownian Motion has the same limiting behavior
as the surface height for flat PNG, namely the $\Af$ process. The
correspondence at the level of top eigenvalues for GOE and the top
layers of the multi-layer flat PNG at a fixed position has been
proven in~\cite{Fer04}, making the conjecture even more plausible.
Knowing the analogue of the Airy process for random growth with
flat initial conditions one can guess the result for $\beta=1$
Dyson's Brownian Motion~\cite{Dys62}.
\begin{conj}\label{Cong2}
The evolution of the largest eigenvalue of $N\times N$ matrices for $\beta=1$ Dyson's Brownian Motion converges, in the limit $N\to\infty$ and properly rescaled, to the process $\Af$.
\end{conj}
Again, the prefactors for the scaling can be easily calculated by matching the known one-point distributions and the behavior of joint distributions at short distances. This conjecture concerns only the largest eigenvalue and with this approach we are unable to make a conjecture for the other eigenvalues.

To make Conjecture~\ref{Cong2} more transparent, we explain it in the simpler case of the two-matrix model. There, one considers two real symmetric $N\times N$ matrices, $M(0)$ and $M(t)$, with joint distribution
\begin{equation}
\frac{1}{Z_{N,t}}\exp\left(-\frac{\Tr(M(0))^2}{2N} \right)\exp\left(-\frac{\Tr(M(t)-q M(0))^2}{2N(1-q^2)}\right)\dx M(0) \dx M(t)
\end{equation}
where $q=\exp(-t/2N)$ and $\dx M(\cdot)=\prod_{1\leq i\leq j \leq
N} \dx M(\cdot)_{i,j}$. Let $\lambda_{\max}(0)$ and
$\lambda_{\max}(t)$ be the largest eigenvalues of $M(0)$ and
$M(t)$. These eigenvalues fluctuate on a scale of order $N^{1/3}$
and are non-trivially correlated if one chooses $t\sim N^{2/3}$.
Then Conjecture~\ref{Cong2} means that, properly rescaled, the
joint distribution of $\lambda_{\max}(0)$ and $\lambda_{\max}(t)$
converges to the two-point joint distribution of the process $\Af$
in the $N\to\infty$ limit.

We also refer to the surveys on the question of universality in mathematics and physics~\cite{De06} and on connections between different models, including random matrices~\cite{FP05}.

\emph{Outline.} The paper is organized as follows: In
Section~\ref{SectModelResult} we state the main result. In
Section~\ref{SectDetMeasure} the kernel of the signed
determinantal point process describing the joint particle
distributions is derived. The kernel involves an orthogonalization
which is carried out in Section~\ref{SectOrtho} for the case of
alternating initial conditions. In Section~\ref{SectAsympt} we
prove the convergence of the properly rescaled kernel to the
kernel $K_{\rm F_1}$. In Appendix~\ref{AppCompactKernel} we
explain how the compact form of the kernel is derived, and in
Appendix~\ref{AppCharlier} we explain how the orthogonalization
can be carried out using classical Charlier orthogonal
polynomials.

\subsection*{Acknowledgment}
P.L. Ferrari would like to thank H. Spohn for useful discussions and for suggesting the compact form of the kernel obtained during our previous work~\cite{FS05b}. A. Borodin was partially supported by the NSF grant DMS-0402047
and the CRDF grant RIM1-2622-ST-04. The work of T. Sasamoto is partly supported by the Grant-in-Aid for Young Scientists (B), the Ministry of Education, Culture, Sports, Science and Technology, Japan.

\section{Model and results}\label{SectModelResult}
In this paper we consider the continuous-time totally asymmetric
simple exclusion process (TASEP) on $\Z$. At any given time $t$, every
site $j\in\Z$ can be occupied at most by one particle. Thus a
configuration of the TASEP can be described by
$\eta=\{\eta_j,j\in\Z|\eta_j\in\{0,1\}\}\in\Omega$. $\eta_j$ is called
the \emph{occupation variable} of site $j$, which is defined by
$\eta_j=1$ if site $j$ is occupied and $\eta_j=0$ if site $j$ is
empty.

The dynamics of the TASEP is defined as follows. Particles jumps on
the neighboring right site with rate $1$ provided that the site is
empty. This means that jumps are independent of each other and are
performed after an exponential waiting time with mean $1$, which
starts from the time instant when the right neighbor site is empty.
More precisely, let $f$: $\Omega\to \R$ be a function depending only
on a finite number of $\eta_j$'s. Then the backward generator of the
TASEP is given by
\begin{equation}\label{1.1}
Lf(\eta)=\sum_{j\in\Z}\eta_j(1-\eta_{j+1})\big(f(\eta^{j,j+1})-f(\eta)\big).
\end{equation}
Here $\eta^{j,j+1}$ denotes the configuration $\eta$ with the
occupations at sites $j$ and \mbox{$j+1$} interchanged. The semigroup
$e^{Lt}$ is well-defined as acting on bounded and continuous functions
on $\Omega$. $e^{Lt}$ is the transition probability of the
TASEP~\cite{Li99}.

\subsubsection*{Joint distributions}
Let us start at time $t=0$ with $N$ particles at positions $y_N <
\ldots < y_2 < y_1$. Then the main result is the joint
distribution of any subset of these particles at time $t>0$. It
turns out that it can be described by a \emph{signed determinantal
point process}, where signed refers to the non-positiveness of the measure.
\begin{thm}\label{ThmJointCorr}
Let $\sigma(1)<\sigma(2)<\ldots<\sigma(m)$ be the indices of $m$ out of the $N$ particles. The joint distribution of their positions $x_{\sigma(k)}(t)$ is given by
\begin{equation}
\Pb\Big(\bigcap_{k=1}^m \big\{x_{\sigma(k)}(t) \geq a_k\big\}\Big)=
\det(\Id-\chi_a K_t\chi_a)_{\ell^2(\{\sigma(1),\ldots,\sigma(m)\}\times\Z)}
\end{equation}
where $\chi_a(\sigma(k),x)=\Id(x<a_k)$. $K_t$ is the extended kernel with
entries
\begin{equation}\label{eqKernelFinal}
K_t(n_1,x_1;n_2,x_2)=-\phi^{(n_1,n_2)}(x_1,x_2)
+\sum_{i=0}^{n_2-1} \Psi^{n_1}_{n_1-n_2+i}(x_1) \Phi^{n_2}_{i}(x_2)
\end{equation}
where
\begin{equation}
\phi^{(n_1,n_2)}(x_1,x_2) = \binom{x_1-x_2-1}{n_2-n_1-1},
\end{equation}
\begin{equation}\label{eq2.4}
\Psi_i^{n}(x)=\frac{1}{2\pi \I} \oint_{\Gamma_0} \frac{\dx w}{w^{i+1}}
\frac{(1-w)^{i}}{w^{x-y_{n-i}}}e^{t(w-1)},
\end{equation}
and the functions $\Phi_i^{n}(x)$, $i=0,\ldots,n-1$, form a family
of polynomials of degree $\le n$ satisfying
\begin{equation}
\sum_{x\in\Z}\Psi_i^{n}(x)\Phi_j^{n}(x)=\delta_{i,j}.
\label{ortho}
\end{equation}
The path $\Gamma_0$ in the definition of $\Psi_i^{n}$ is any simple
loop, anticlockwise oriented, which includes the pole at $w=0$ but
\emph{not} the one at $w=1$.
\end{thm}
The dependence on the set $\{y_i\}$ is hidden in the definition of the $\Phi_i^n$'s
and the $\Psi_i^n$'s but is omitted, since the set $\{y_i\}$ is fixed
in the following.

\subsubsection*{Alternating initial configuration}
Now we consider alternating initial configuration, namely
\begin{equation}
\eta_i(0)=\left\{\begin{array}{ll} 1,&\textrm{if }i\textrm{ is
even},\\ 0,&\textrm{if }i\textrm{ is odd}.\end{array}\right.
\end{equation}
The alternating initial configuration can be obtained by taking
$2N$ particles around the origin, for example at positions
$2\Z\cap[-2N,2N-2]$, and then taking the limit $N\to\infty$. In
Lemma~\ref{lemOrtho} we do the orthogonalization, i.e., construct
$\Phi_i^n$'s which satisfy (\ref{ortho}) for this special case,
from which the kernel $K_t$ is obtained.
\begin{thm}\label{ThmKernel} Let particle with label $n_i$ start
at $-2i$, $i\in \Z$. At time $t$, the particles are at positions
$x_i$. The kernel (\ref{eqKernelFinal}) for the alternating
initial configuration is given by
\begin{equation}\label{eqKernelD2}
K_t(n_1,x_1;n_2,x_2)=-\binom{x_1-x_2-1}{n_2-n_1-1}
+\frac{-1}{2\pi \I}\oint_{\Gamma_0} \dx v \frac{(1+v)^{x_2+n_1+n_2}}{(-v)^{x_1+n_1+n_2+1}}e^{-t(1+2v)}
\end{equation}
where $\Gamma_0$ is any simple loop, anticlockwise oriented, which includes the pole at $v=0$ but do not include $v=-1$.
\end{thm}

\subsubsection*{Scaling limit}
The particle density is $1/2$ and since particles jump to the
right with rate $1$ provided the site is empty, the mean speed of
the particles is $1/2$. Let us number the particles from right to
left with $y_1(0)=0$ as reference point, i.e., $y_i(0)=-2(i-1)$,
$i\in\Z$. Then the particles which at time $t$ are close to $x=0$
are the particles with numbers close to $t/4$. From universality
we know also that the scaling exponent for fluctuations should be
$1/3$ and the one for spatial correlations should be $2/3$.
Therefore, the scaling limit to be considered is
\begin{equation}\label{Rescaling}
x_i=-2 u_i t^{2/3}- s_i t^{1/3},\quad n_i=t/4+ u_i t^{2/3}.
\end{equation}
\textbf{Remark}: The scaling exponents for this model are
determined by the requirement that there is a non-trivial limit.
The numerical factor in front of $t^{1/3}$ is chosen so that the
single-time kernel has a simple form, $\Ai(x+y)$. The numerical
factors for the $t^{2/3}$ terms are set in such a way that the
propagator in (\ref{eqKernelCompact}) is generated by the
Laplacian without additional prefactors.
Universality argument is not needed to obtain the result, but it is useful to predict the correct answer.

In Section~\ref{SectAsympt} we carry out the asymptotic analysis for
the pointwise convergence of the kernel, with the following result.
\begin{thm}[Pointwise convergence of the kernel]\label{ThmConvPt}
  Let $x_1,n_1,x_2,n_2$ be rescaled as in (\ref{Rescaling}). Then, for any $s_1,s_2,u_1,u_2\in\R$ fixed,
\begin{equation}\label{KernelAsympt}
\lim_{t\to\infty}K_t(n_1,x_1;n_2,x_2) t^{1/3}2^{x_2-x_1} = K_{\rm F_1}(u_1,s_1;u_2,s_2)
\end{equation}
where the extended kernel $K_{\rm F_1}$ is given in (\ref{eqKernelExpanded}).
\end{thm}

In this paper we do not perform the asymptotic analysis necessary to get convergence of the Fredholm determinants. We will do the complete analysis in a discrete-time version of the TASEP in a forthcoming paper~\cite{BFP06}, from which the continuous time limit follows as a corollary. Nevertheless, it is instructive for the reader to see the implications of the convergence of the Fredholm determinant.

Let $\Af$ be the process with $m$-point joint distributions at
\mbox{$u_1< u_2< \ldots < u_m$} given by
\begin{equation}
\Pb\Big(\bigcap_{k=1}^m\{\Af(u_k)\leq s_k\}\Big)=
\det(\Id-\chi_s K_{\rm F_1}\chi_s)_{L^2(\{u_1,\ldots,u_m\}\times \R)}
\end{equation}
where $\chi_s(u_k,x)=\Id(x>s_k)$. The convergence of Fredholm
determinant would then imply
\begin{equation}
\lim_{t\to\infty}\frac{x_{[t/4+ut^{2/3}]}(t)+2ut^{2/3}}{-t^{1/3}}=\Af(u),
\end{equation}
with the convergence understood in the sense of finite-dimensional distributions.

\section{Signed determinantal point process}\label{SectDetMeasure}
In this section we prove Theorem~\ref{ThmJointCorr}. Consider the
TASEP with $N$ particles starting at time $t=0$ at positions $y_N <
\ldots < y_2 < y_1$. The first step is to obtain the probability that
at time $t$ these particles are at positions $x_N < \ldots < x_2 <
x_1$, which we denote by
\begin{equation}
G(x_1,\ldots,x_N;t)=\Pb((x_N,\ldots,x_1;t)|(y_N,\ldots,y_1;0)).
\end{equation}
This function has been determined before using Bethe-Ansatz method~\cite{Sch97}.
\begin{lem}[Sch\"utz \cite{Sch97}]\label{lem1}
  The transition probability has a determinantal form
\begin{equation}\label{eqGreen}
G(x_1,\ldots,x_N;t)=\det(F_{i-j}(x_{N+1-i}-y_{N+1-j},t))_{1\leq i,j\leq N}
\end{equation}
with
\begin{equation}\label{eqFn}
F_{n}(x,t)=\frac{(-1)^n}{2\pi \I} \oint_{\Gamma_{0,1}} \frac{\dx w}{w}
\frac{(1-w)^{-n}}{w^{x-n}}e^{t(w-1)},
\end{equation}
where $\Gamma_{0,1}$ is any simple loop oriented anticlockwise which
includes $w=0$ and $w=1$.
\end{lem}

This representation of the transition probability was utilized to
study the current fluctuations in \cite{NS04,RS05}. To study the joint distribution, we need a decomposition of
$G(x_1,\ldots,x_N;t)$ given in the next lemma. This decomposition
is obtained using only the recurrence relation
\begin{equation}\label{eqRel2}
F_{n-1}(x,t)=F_n(x,t)-F_n(x+1,t)
\end{equation}
and its integrated form
\begin{equation}\label{RecRel}
F_{n+1}(x,t)=\sum_{y\geq x} F_n(y,t).
\end{equation}
Actually, (\ref{RecRel}) comes from (\ref{eqRel2}) and the fact that
$\lim_{y\to\infty}F_n(y,t)=0$ fast enough.  The other property needed
to obtain Theorem~\ref{ThmJointCorr} is the following.
Comparing (\ref{eq2.4}) and (\ref{eqFn}), we have
\begin{equation}
\Psi_k^N(x)=(-1)^k F_{-k}(x-y_{N-k},t)
\label{PsiF}
\end{equation}
for $k\geq 0$.
Notice that, for $n=-k<0$, (\ref{eqFn}) defining $F_n$ has actually
only one pole at $w=0$. We then get the relation
\begin{equation}\label{eqRel3}
F_{n+1}(x,t)=-\sum_{y<x}F_{n}(y,t),
\end{equation}
which translates into
\begin{equation}\label{eqRel4}
\Psi^N_{N-k}(x)=\sum_{y<x}\Psi^{N+1}_{N+1-k}(y).
\end{equation}
In the definition of the $\Psi^N_k$'s in Theorem~\ref{ThmJointCorr},
the path $\Gamma_0$ includes only the pole at the origin, exactly
because we need (\ref{eqRel4}) to hold also for $k<0$.

\begin{lem}\label{lem2}
  Let us denote $x_k=x_1^k$, $k=1,\ldots,N$. Then
\begin{equation}\label{eqLem2}
G(x_1,\ldots,x_N;t)=\sum_{\cal D}
\det(F_{-j}(x_{i+1}^N-y_{N-j},t))_{0\leq i,j \leq N-1}
\end{equation}
where the sum is over the following set
\begin{equation}
{\cal D}=\{x_i^j,2 \leq i \leq j \leq N | x_i^j>x_i^{j+1},x_i^j\geq x_{i-1}^{j-1}\}.
\end{equation}
See Figure~\ref{figScheme} for a graphical representation of $\cal D$.
\begin{figure}[t!]
\begin{center}
  \psfrag{m}[c]{$<$} \psfrag{e}[c]{$\leq$} \psfrag{x11}{$x_1^1$}
  \psfrag{x12}{$x_1^2$} \psfrag{x13}{$x_1^3$} \psfrag{x14}{$x_1^4$}
  \psfrag{x22}{$x_2^2$} \psfrag{x23}{$x_2^3$} \psfrag{x24}{$x_2^4$}
  \psfrag{x33}{$x_3^3$} \psfrag{x34}{$x_3^4$} \psfrag{x44}{$x_4^4$}
  \includegraphics[height=3.5cm]{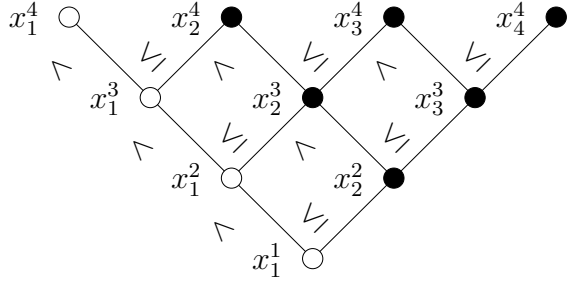}
\caption{Graphical representation of the domain of integration $\cal D$ for $N=4$. One has to ``integrate'' out the variables $x_i^j$, $i\geq 2$ (i.e., the black dots). The positions of $x_1^k$, $k=1,\ldots,N$ are given (i.e., the white dots).}
\label{figScheme}
\end{center}
\end{figure}
\end{lem}
This lemma is actually more general as shown in~\cite{Sas05}. By
applying just the recursion relation (\ref{RecRel}) the domain of
summation in Lemma~\ref{lem2} would be ${\cal D'}=\{x_i^j,2 \leq i
\leq j \leq N | x_i^j\geq x_{i-1}^{j-1}\}$ instead of $\cal D$.
The reduction of the summation domain to $\cal D$ uses only the
antisymmetry of the determinant. Thus the same holds for any
antisymmetric function $f$. It might be interesting in other
applications, so we state it explicitly.
\begin{lem} Let $f$ an antisymmetric function of $\{x_1^{N},\ldots,x_N^{N}\}$. Then, whenever $f$ has enough decay to make the sums finite,
\begin{equation}
\sum_{\cal D}f(x_1^{N},\ldots,x_N^{N})=\sum_{\cal D'}f(x_1^{N},\ldots,x_N^{N})
\end{equation}
where
\begin{eqnarray}
{\cal D}&=&\{x_i^j,2 \leq i \leq j \leq N | x_i^j>x_i^{j+1},x_i^j\geq x_{i-1}^{j-1}\},\nonumber \\
{\cal D'}&=&\{x_i^j,2 \leq i \leq j \leq N | x_i^j\geq x_{i-1}^{j-1}\},
\end{eqnarray}
and the positions $x_1^1>x_1^2>\ldots>x_1^N$ being fixed.
\end{lem}

\begin{proofOF}{Lemma~\ref{lem2}}
  The proof consists in applying the property (\ref{RecRel})
  iteratively and using the multilinearity of the determinants. From
  Lemma~\ref{lem1} we have
\begin{equation}\label{eq27}
G(x_1^1,\ldots,x_1^N;t)=\det\left[\begin{array}{ccc} F_0(x_1^N-y_N,t)&
\cdots & F_{-N+1}(x_1^N-y_1,t) \\ \vdots & \ddots & \vdots \\
F_{N-1}(x_1^1-y_N,t)&\cdots & F_0(x_1^1-y_1,t)\end{array}\right].
\end{equation}
The first step is to rewrite the last row as
\begin{equation}
\sum_{x_2^2 \geq x_1^1}\left[\begin{array}{ccc}
F_{N-2}(x_2^2-y_N,t)&\cdots & F_{-1}(x_2^2-y_1,t)\end{array}\right].
\end{equation}
The second step is to apply the same procedure to the last and
second to last rows, which become
\begin{equation}
\sum_{x_2^2 \geq x_1^1}\sum_{x_3^3\geq x_2^2} \left[\begin{array}{ccc}
F_{N-3}(x_3^3-y_N,t)&\cdots & F_{-2}(x_3^3-y_1,t)\end{array}\right]
\end{equation}
and
\begin{equation}
\sum_{x_2^3 \geq x_1^2} \left[\begin{array}{ccc}
F_{N-3}(x_2^3-y_N,t)&\cdots & F_{-2}(x_2^3-y_1,t)\end{array}\right]
\end{equation}
respectively. At this point we have
\begin{equation}
(\ref{eq27})=\sum_{x_2^2 \geq x_1^1}\sum_{x_3^3\geq x_2^2}\sum_{x_2^3
\geq x_1^2} \det\left[\begin{array}{ccc} F_0(x_1^N-y_N,t)& \cdots &
F_{-N+1}(x_1^N-y_1,t) \\ \vdots & \ddots & \vdots \\
F_{N-3}(x_1^3-y_N,t)&\cdots & F_{-2}(x_1^3-y_1,t) \\
F_{N-3}(x_2^3-y_N,t)&\cdots & F_{-2}(x_2^3-y_1,t) \\
F_{N-3}(x_3^3-y_N,t)&\cdots & F_{-2}(x_3^3-y_1,t) \end{array}\right].
\end{equation}
The determinant is antisymmetric in the variables $(x_2^3,x_3^3)$,
therefore the contribution of the symmetric part of the summation
domain of $\sum_{x_3^3\geq x_2^2}\sum_{x_2^3 \geq x_1^2}$ is zero.
Since $\sum_{x_3^3\geq x_2^2}\sum_{x_2^3 \geq x_1^2}=\sum_{x_3^3\geq
  x_2^2}\sum_{x_2^3 \in[x_1^2,x_2^2)}+\sum_{x_3^3\geq
  x_2^2}\sum_{x_2^3\geq x_2^2}$, the symmetric part of the domain is
$\{x_3^3\geq x_2^2,x_2^3\geq x_2^2\}$, thus the contribution coming
from the last sum is zero.

We iterate the same procedure. More precisely, for $k=4,\ldots,N$,
we apply (\ref{RecRel}) to the last $(k-1)$ rows. The new summing
variable for the last row is denoted by $x_{k}^k$, the second last
row $x_{k-1}^k$, and so on. Finally, we can delete the sums over
the symmetric domain in $(x_2^k,\ldots,x_k^k)$. In this way we get
the result
\begin{equation}\label{3.14}
G(x_1^1,\ldots,x_1^N;t)=\sum_{\cal D}\det\left[\begin{array}{ccc}
F_0(x_1^N-y_N,t)& \cdots & F_{-N+1}(x_1^N-y_1,t) \\ \vdots & \ddots &
\vdots \\ F_{0}(x_N^N-y_N,t)&\cdots &
F_{-N+1}(x_N^N-y_1,t)\end{array}\right].
\end{equation}
\end{proofOF}

\begin{figure}[t!]
\begin{center}
  \psfrag{0}[r]{$0$} \psfrag{1}[r]{$1$} \psfrag{2}[r]{$2$}
  \psfrag{3}[r]{$3$} \psfrag{4}[r]{$N=4$} \psfrag{N}[r]{$N+1$}
  \psfrag{j}[c]{$x$} \psfrag{M}[c]{$U$} \psfrag{n}[c]{$n$}
  \psfrag{00}[c]{$0$} \psfrag{y1}[c]{$y_1$} \psfrag{y2}[c]{$y_2$}
  \psfrag{y3}[c]{$y_3$} \psfrag{y4}[c]{$y_4$} \psfrag{y5}[c]{$y_5$}
  \includegraphics[height=4cm]{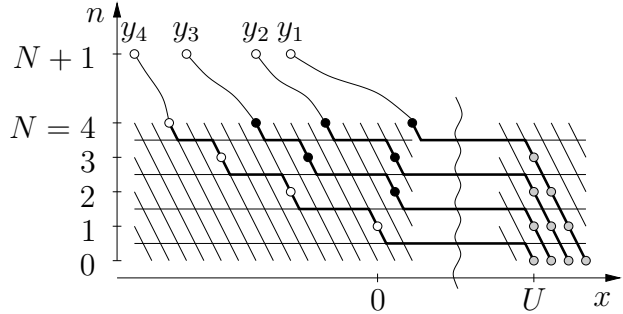}
\caption{LGV scheme for $N=4$. The LGV graph is left/up-left directed with weight $1$ for each edge. From $N$ to $N+1$ the transitions are the $F$'s.}
\label{figSchemeLGV}
\end{center}
\end{figure}

This is the decomposition used in~\cite{Sas05}. The integrations
variables \mbox{$\{x_i^n,i=1,\ldots,n\}$} can be interpreted as the
positions of particles labelled by $i=1,\ldots,n$ at time $n$. For
example, $x_1^1,\ldots,x_1^n$ is the trajectory of particle $1$,
see also Figure~\ref{figSchemeLGV}. This is just a mathematical
construction which should not to be confused with the real TASEP
particles and the natural time in the TASEP positions, which at
this stage is just the fixed parameter $t$. At time $n$ there are
$n$ particles at positions $x_1^n,\ldots,x_n^n$. At time $n+1$,
they jump to a randomly uniformly chosen position satisfying
$x_k^{n+1}\in [x_{k-1}^n,x_k^n)$ with the $(n+1)$st particle added
at position $x_{n+1}^{n+1} (\geq x_n^n)$. Then the weight of a
configurations of $x_i^n$'s is given by
\begin{eqnarray}\label{eqWeightA}
& &W(\{x_i^n; 1\leq i \leq n \leq N\})\\ & &=\Big(\prod_{n=2}^N
\det(\Id(x_i^{n-1}>x_j^{n}))_{1\leq i,j \leq n}\Big)
\det(F_{-j}(x^N_{i+1}-y_{N-j},t))_{0\leq i,j \leq N-1}, \nonumber
\end{eqnarray}
where we set $x_n^{n-1}=\infty$. The products of determinants in
(\ref{eqWeightA}) might look complicated. However, one can verify
that whenever some of the $x_i$'s do not satisfy the inequalities
of $\cal D$, then at least one of the determinant vanishes. On the
other hand, if the set of $x_i$'s belongs to $\cal D$, then each
determinant has value $1$.

The form of the weight suggests that the correlation functions could be determinantal. It is like to have a Lindstr\"om-Gessel-Viennot (LGV) scheme~\cite{Vie77}, see~\cite{Ste90} for a nice exposition, with a sort of reservoir of particles at $\infty$ and at each time-step a new particle is introduced. The LGV scheme is a sort of generalization on a class of directed graphs of the Karlin-McGregor result for diffusions~\cite{KM59}. Determinantal form of correlation functions appeared in different contexts~\cite{EM97,FNH99,OR01,RB04,Jo03b}. Although we do not use the LGV scheme in the proof, it might be interesting for the reader to see how the weight (\ref{eqWeightA}) can be described in this framework. The situation given by the weight (\ref{eqWeightA}) corresponds to the limit $U\to\infty$ of the system with fixed number of particles illustrated in Figure~\ref{figSchemeLGV}. As $U\to\infty$, the extra particles are not seen, they goes to $\infty$ in a sort of reservoir.

The proof of Theorem~\ref{ThmJointCorr} is an application of the following Lemma, which is proven by using the framework of~\cite{RB04}.
\begin{lem}\label{SasLemma}
Assume we have a signed measure on $\{x_i^n,n=1,\ldots,N,i=1,\ldots,n\}$ given in the form,
\begin{equation}\label{Sasweight}
 \frac{1}{Z_N}\prod_{n=1}^{N-1} \det[\phi_n(x_i^n,x_j^{n+1})]_{1\leq i,j\leq n+1} \det[\Psi_{N-i}^{N}(x_{j}^N)]_{1\leq i,j \leq N},
\end{equation}
where $x_{n+1}^n$ are some ``virtual'' variables and $Z_N$ is a normalization constant. If $Z_N\neq 0$, then the correlation functions are determinantal.

To write down the kernel we need to introduce some notations. Define
\begin{equation}\label{Sasdef phi12}
\phi^{(n_1,n_2)}(x,y)=\left\{\begin{array}{ll}(\phi_{n_1}* \cdots *
\phi_{n_2-1})(x,y),& n_1<n_2,\\ 0,& n_1\geq n_2,\end{array}\right.
\end{equation}
where $(a* b)(x,y)=\sum_{z\in\Z}a(x,z) b(z,y)$, and, for $1\leq n<N$,
\begin{equation}\label{Sasdef_psi}
\Psi_{n-j}^{n}(x) := (\phi^{n,N} * \Psi_{N-j}^{N})(y), \quad j=1,2,\ldots,N.
\end{equation}
Set $\phi_0(x_1^0,x)=1$. Then the functions
\begin{equation}
\{ (\phi_0*\phi^{(1,n)})(x_1^0,x), \dots,(\phi_{n-2}*\phi^{(n-1,n)})(x_{n-1}^{n-2},x), \phi_{n-1}(x_{n}^{n-1},x)\}
\end{equation}
are linearly independent and generate the $n$-dimensional space $V_n$. Define a set of functions $\{\Phi_j^{n}(x), j=0,\ldots,n-1\}$ spanning $V_n$ defined by the orthogonality relations
\begin{equation}\label{Sasortho}
\sum_x \Phi_i^n(x) \Psi_j^n(x) = \delta_{i,j}
\end{equation}
for $0\leq i,j\leq n-1$.

Under Assumption (A): $\phi_n(x_{n+1}^n,x)=c_n \Phi_0^{(n+1)}(x)$, for some $c_n\neq 0$, $n=1,\ldots,N-1$, the kernel takes the simple form
\begin{equation}\label{SasK}
K(n_1,x_1;n_2,x_2)= -\phi^{(n_1,n_2)}(x_1,x_2)+ \sum_{k=1}^{n_2} \Psi_{n_1-k}^{n_1}(x_1) \Phi_{n_2-k}^{n_2}(x_2).
\end{equation}
\end{lem}
\textbf{Remarks:} Without Assumption (A), the correlations functions are still determinantal but the formula is modified as follows. Let $M$ be the $N\times N$ dimensional matrix defined by $[M]_{i,j}=(\phi_{i-1}*\phi^{(i,N)}*\Psi^N_{N-j})(x_i^{i-1})$. Then
\begin{eqnarray}
& &K(n_1,x_1;n_2,x_2)\\
&=& -\phi^{(n_1,n_2)}(x_1,x_2)+ \sum_{k=1}^{n_2} \Psi_{n_1-k}^{n_1}(x_1) \sum_{l=1}^N [M^{-1}]_{k,l} (\phi_{l-1}*\phi^{(l,n_2)})(x_l^{l-1},x_2).\nonumber
\end{eqnarray}

The analogue of the determinantal representation (\ref{eqGreen}) for particle-dependent hopping rates has been recently obtained~\cite{RS06}. Lemma~\ref{SasLemma} might be applied in this context too.

\begin{proofOF}{Lemma~\ref{SasLemma}}
We apply Proposition 1.2 of~\cite{RB04} and we try to stick as much as possible to the notations therein. Let, for $n=1,\ldots,N$, $\Xf^{(n)}$ denote the space of $\{x_i^n,i=1,\ldots,n\}$, $\Yf=\Xf^{(1)}\cup\ldots\cup\Xf^{(N)}$, and let \mbox{$\Xf=\{x_1^0,x_2^1,\ldots,x_N^{N-1}\}\cup\Yf$} be the space on which our measure (\ref{Sasweight}) is defined. Let $T^{(n,n+1)}$ be the matrix with entries
\begin{equation}\label{SasmatrixT}
[T^{(n,m)}]_{i,j}=\phi^{(n,n+1)}(x_i^n,x_j^{n+1}),\quad 1\leq i,j\leq n+1
\end{equation}
and
\begin{equation}\label{SasMatrixPsi}
[\Psi^{(N)}]_{i,j}=\Psi^N_{N-j}(x_i^N),\quad  1\leq i,j \leq N.
\end{equation}
Then the weight (\ref{Sasweight}) is proportional to the determinant of
\begin{equation}
\left[\begin{array}{ccccc}
0 & -T^{(1,2)} & 0 & \cdots & 0 \\
0 & 0 & -T^{(2,3)} & \cdots & 0 \\
\vdots & \vdots & \vdots & \ddots & \vdots \\
0 & 0 & 0 & \cdots & -T^{(N-1,N)} \\
\Psi^{(N)} & 0 & 0 & 0 & 0
\end{array}
\right].
\end{equation}

We are interested in the measure on $\Yf$ only, thus we change the ordering by putting the variables $x_1^0,\ldots,x_N^{N-1}$ at the beginning. Let us define the $n\times (n+1)$ matrix $W_{[n,n+1)}$ by
\begin{equation}\label{SasmatrixW}
[W_{[n,n+1)}]_{i,j}=\phi^{(n,n+1)}(x_i^n,x_j^{n+1}),\quad 1\leq i\leq n,1\le j \leq n+1,
\end{equation}
and the $N\times (m+1)$ matrix $E_m$ by
\begin{equation}\label{SasmatrixE}
[E_{m}]_{i,j}=\left\{\begin{array}{ll}
\phi_m(x_{m+1}^m,x_j^{m+1}),&i=m+1,1\leq j \leq m+1,\\
0,&\textrm{otherwise}.
\end{array}\right.
\end{equation}
Then the weight (\ref{Sasweight}) is proportional to a
suitable symmetric minor of $L$, with
\begin{equation}\label{SasMatrixL}
L=\left[\begin{array}{cccccc}
0 & E_0 & E_1 & E_2 &\ldots & E_{N-1} \\
0 & 0 & -W_{[1,2)} & 0 & \cdots & 0 \\
0 & 0 & 0 & -W_{[2,3)} & \ddots & \vdots \\
\vdots & \vdots & \vdots & \ddots & \ddots &0    \\
0 & 0 & 0 & 0 &\cdots & -W_{[N-1,N)} \\
\Psi^{(N)} & 0 & 0 & 0 & \cdots & 0
\end{array}
\right].
\end{equation}

By Proposition 1.2 of~\cite{RB04}, the point-measure on $\Yf$ is determinantal with correlation kernel given by
\begin{equation}
K=\Id_{\Yf}-(\Id_\Yf+L)^{-1}\Big|_{\Yf\times\Yf}
\end{equation}
provided that the partition function $Z_N\neq 0$. With the decomposition of $\Xf=\{x_1^0,x_2^1,\ldots,x_N^{N-1}\}\cup\Yf$, we have a block decomposition of $L$ as
\begin{equation}
L=\left[
    \begin{array}{cc}
      0 & B \\
      C & D_0 \\
    \end{array}
  \right]
\end{equation}
with $B=[E_0,\ldots,E_{N-1}]$, $C=[0,\ldots,0,\Psi^{(N)}]^t$, and $D_0$ equal to $L$ without the first line and column of the block representation (\ref{SasMatrixL}). Let $D=\Id+D_0$, then (see, e.g., Lemma 1.5 of~\cite{RB04}) the kernel is given by
\begin{equation}
K=\Id-D^{-1}+D^{-1} C M^{-1} B D^{-1},\quad M=B D^{-1} C.
\end{equation}
$D^{-1}$ was already computed in Lemma 1.5 of~\cite{RB04}, with the result
\begin{equation}
D^{-1}=\left[\begin{array}{cccc}
\Id & W_{[1,2)} & \cdots & W_{[1,N)} \\
0& \Id  & \ddots & \vdots \\
\vdots & \ddots  & \ddots & W_{[N-1,N)} \\
0& 0 &  0 &\Id
\end{array}
\right],
\end{equation}
where
\begin{equation}
W_{[n,m)}=\left\{
\begin{array}{cc}
W_{[n,n+1)}\cdots W_{[m-1,m)}, & m>n, \\
0, & m\leq n.
\end{array}
\right.
\end{equation}
Thus the $(n,m)$ block of $\Id-D^{-1}$ is $W_{[n,m)}$. Next, we have
\begin{equation}
D^{-1} C = \left[\begin{array}{c}
W_{[1,N)} \Psi^{(N)} \\
\vdots \\
W_{[N-1,N)} \Psi^{(N)} \\
\Psi^{N}
\end{array}\right],
\end{equation}
and
\begin{equation}
B D^{-1} = \left[\begin{array}{cccc}
E_0 & E_0 W_{[1,2)}+E_1 & \cdots & \sum_{k=1}^{N-1} E_{k-1} W_{[k,N)}+ E_{N-1}
\end{array}\right].
\end{equation}
Therefore the $(n,m)$ block of the correlation kernel is given by
\begin{equation}\label{Sas19}
K^{(n,m)}=-W_{[n,m)}+W_{[n,N)} \Psi^{(N)} M^{-1} \Big(\sum_{k=1}^{m-1} E_{k-1} W_{[k,m)}+ E_{m-1}\Big).
\end{equation}
Using (\ref{Sasdef phi12}) one gets $[W_{[n,m)}]_{i,j}=\phi^{(n,m)}(x_i^n,x_j^m)$. Moreover by (\ref{Sasdef_psi}) we have
\begin{equation}\label{Sas22}
[W_{[n,N)}\Psi^{(N)}]_{i,j}=\sum_{y}\phi^{(n,m)}(x_i^n,y)\Psi^N_{N-j}(y)=\Psi^{n}_{n-j}(x_i^n).
\end{equation}
It remains to evaluate the last part of (\ref{Sas19}). For the following $N\times m$ matrix we have
\begin{equation}\label{Sas23}
\Big[\sum_{k=1}^{m-1} E_{k-1} W_{[k,m)}+ E_{m-1}\Big]_{i,j}=\left\{\begin{array}{ll}
(\phi_{i-1}*\phi^{(i,m)})(x_i^{i-1},x_j^m),& 1\leq i \leq m,\\
0, & m+1\leq i \leq N.
\end{array}\right.
\end{equation}
Notice that the functions in (\ref{Sas23}) form a basis of $V_m$. Thus we can define a $m\times m$ matrix $B_m$ which does a change of basis to $\{\Phi^m_{m-1}(x),\ldots,\Phi^m_0(x)\}$, namely
\begin{equation}
(\phi_{i-1}*\phi^{(i,m)})(x_i^{i-1},x)=\sum_{l=1}^m [B_m]_{i,l} \Phi^m_{m-l}(x).
\end{equation}
We multiply this equation by $\sum_x \Psi^m_{m-j}(x)$ and obtain
\begin{equation}
[B_m]_{i,j}=(\phi_{i-1}*\phi^{(i,m)}*\Psi^m_{m-j})(x_i^{i-1}).
\end{equation}
In particular, we have $B_N=M$. Let us define the $N\times m$ matrix
\begin{equation}\label{SasMatrixPhi}
[\Phi^{(m)}]_{i,j}=\left\{\begin{array}{ll} \Phi^m_{m-i}(x_j^m),& 1\leq i\leq m,\\
0,&m+1\leq i \leq N.\end{array}\right.
\end{equation}
Then
\begin{equation}\label{Sas25}
M^{-1} \Big(\sum_{k=1}^{m-1} E_{k-1} W_{[k,m)}+ E_{m-1}\Big) = B_N^{-1} \left[
\begin{array}{cc}
B_m & 0 \\
0 & 0 \\
\end{array}
\right] \Phi^{(m)}.
\end{equation}

Assume the condition (B): $(\ref{Sas25})=\Phi^{(m)}$ for $m=1,\ldots,N$. Then we get the simple form of the kernel, (\ref{SasK}), of the Lemma. However, this is not always the case. For $1\leq i,j\leq m$, we obtain using (\ref{Sasdef_psi}),
\begin{equation}\label{Sas27}
[B_m]_{i,j}=(\phi_{i-1}*\phi^{(i,m)}*\Psi^m_{m-j})(x_i^{i-1})=(\phi_{i-1}*\phi^{(i,N)}*\Psi^N_{N-j})(x_i^{i-1})=[B_N]_{i,j}.
\end{equation}
Thus we can write $B_N=\left[\begin{array}{cc}
B_m & \star \\
Q_m & \star \\
\end{array}\right]$ for some $(N-m)\times m$ matrix $Q_m$. By multiplying on both sides by $B_N$ the condition (B), we see that (B) is equivalent to
\begin{equation}
\sum_{k=1}^m [Q_m]_{i,k} \Phi^{m}_{m-k}(x)=0
\end{equation}
for all $x$ and for all $i=1,\ldots,N-m$, and for all $m=1,\ldots,N$. But the functions $\Phi^m_{m-k}(x)$ form a basis of $V_m$, thus (B) is fulfilled iff $Q_m=0$ for all $m=1,\ldots,N$. Thus (B) is equivalent to the condition $B_N$ is an upper-diagonal matrix.

Assume $B_m$ upper diagonal for some $m$. This is verified for $m=1$ where $B_1=1$. Then by (\ref{Sas27})
$[B_{m+1}]_{i,j}=[B_m]_{i,j}$ for $i=1,\ldots,m$, and $B_{m+1}$ is still upper-diagonal iff
\begin{equation}
[B_{m+1}]_{m+1,j}\equiv (\phi_m*\Psi^{m+1}_{m+1-j})(x_{m+1}^m)= c_m \delta_{j,m+1}, \quad c_m\neq 0.
\end{equation}
$c_m\neq 0$ because $Z_N\neq 0$. Finally, the orthogonal relations (\ref{Sasortho}) imply that
$\phi_m(x_{m+1}^m,x)=c_m \Phi_0^{m+1}(x)$, which is Assumption (A) of the Lemma.
\end{proofOF}

\begin{proofOF}{Theorem~\ref{ThmJointCorr}}
It consists in an application of Lemma~\ref{SasLemma}, with
\begin{equation}
\phi_{n}(x_i^{n},x_j^{n+1})= \Id(x_i^{n}>x_j^{n+1}),\quad n=1,\ldots,N-1,
\end{equation}
and
\begin{equation}
\Psi^{N}_{N-i}(x)=(-1)^{N-i} F_{-N+i}(x-y_{i},t),\quad i=1,\ldots,N.
\end{equation}
An important point is that in (\ref{eqFn}) the functions $F_i$'s are
defined by an integral enclosing $w=0$ and $w=1$. At this stage, we
have only functions $F_i$'s for $i\leq 0$. In this case, as mentioned
around (\ref{PsiF}), $w=1$ is actually not a pole, thus the weight (\ref{eqWeightA}) and the weight (\ref{Sasweight}) with the above replacements are proportional. The definition of the $\Psi$'s using only one pole fit exactly in the framework of Lemma~\ref{SasLemma}. In fact, by (\ref{eqRel4}), we have the composition rule
\begin{equation}\label{eq3.25a}
(\phi*\Psi^{n+1}_{n+1-j})(x)=\Psi^n_{n-j}(x),
\end{equation}
which gives (\ref{Sasdef_psi}) by iterations.
In our setting, if we sum up all the variables $\{x_j^m,1\leq m <n,1\leq j \leq m\}$, we get a Vandermonde determinant in the variables $x_j^n$'s. Thus the space $V_n$ of Lemma~\ref{SasLemma} is generated by $\{1,x,\ldots,x^{n-1}\}$ and $\Phi^n_k$ are polynomials of order at most $n-1$. A simple computation using (\ref{eq2.4}) leads to
\begin{equation}
\sum_x \Psi_{j}^{n}(x) = \begin{cases}0, & j=1,\ldots,n-1, \\ 1, & j=0,\end{cases}
\end{equation}
which, together with (\ref{Sasortho}) leads to $\Phi^n_0(x)=1=\phi_{n-1}(\infty,x)$. Thus we have a determinantal system with kernel (\ref{SasK}), which can be rewritten as
\begin{equation}
K_t(n_1,x_1;n_2,x_2)=-\phi^{(n_1,n_2)}(x_1,x_2)+\sum_{i=0}^{n_2-1}
\Psi^{n_1}_{n_1-n_2+i}(x_1) \Phi^{n_2}_{i}(x_2).
\end{equation}

Since in this paper we explain the detail of the derivation in~\cite{Sas05}, it is useful to point out a difference. There one does not obtain directly $\phi^{(n_1,n_2)}$ as in Theorem~\ref{ThmJointCorr}. There the one-time transition becomes \mbox{$\phi_n(x,y)=-\Id(y\geq x)$} and the representation (\ref{eqFn}) has to be used instead. The final form of the kernel (eq.\ (13) in~\cite{Sas05}) comes from splitting the contribution from the pole at $w=1$ and the remainder.
In the geometric picture, it corresponds to have the conjugate LGV graph with reservoir of particles at $-\infty$ instead of at $+\infty$.
\end{proofOF}

\section{Orthogonalization}\label{SectOrtho}
In order to get the kernel for the alternating initial
configuration, i.e., for the case where particles initially occupy
exactly the sublattice $2\Z$, we start with a finite number of
particles, $2N$. In the second step we will focus on the region
where the $N$-dependence vanishes. In this way we will get the
kernel for the system we actually are interested in.
\begin{figure}[t!]
\begin{center}
  \psfrag{0}[c]{$0$} \psfrag{x}[l]{$x$} \psfrag{t}[l]{$t$}
  \psfrag{n}[c]{$-2N$} \psfrag{m}[c]{$2N-2$}
  \includegraphics[height=4cm]{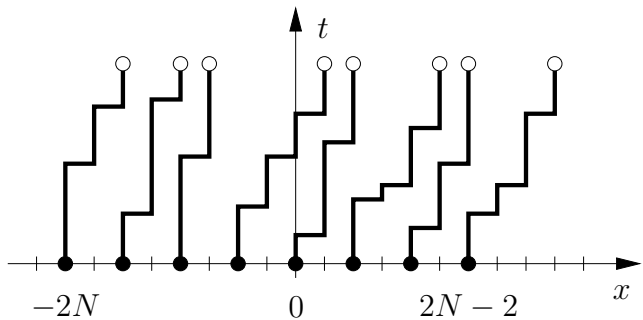}
\caption{Trajectories of the $2N$ particles. The black dots are the initial positions and the white dots are the positions of the particles at some later time $t$. This is a scheme leading, in the $N\to\infty$ limit, the alternating initial configuration on $\Z$.}\label{figInitialCond}
\end{center}
\end{figure}

Consider the case where at time $t=0$ there are $2N$ particles placed
every second site centered around the origin, see
Figure~\ref{figInitialCond}, namely
\begin{equation}
y_i=2N-2i,\quad i=1,\ldots,2N.
\end{equation}
From Theorem~\ref{ThmJointCorr}, the kernel is known once the
orthogonalization is carried out. Here we state the result and a
short proof. In Appendix~\ref{AppCharlier} we explain a
constructive way of obtaining needed functions using Charlier
orthogonal polynomials.
\begin{lem}\label{lemOrtho} The functions $\Psi^{n}_k(x)$ and $\Phi^{n}_k(x)$
have the following integral representations. Let $z=x+2n-2N$. Then
\begin{equation}\label{eqFctPsi}
\Psi^n_k(x)=\frac{(-1)^k}{2\pi \I}\oint_{\Gamma_0}\frac{\dx w}{w^{z+1}} e^{(w-1)t} ((w-1)w)^k
\end{equation}
and
\begin{equation}\label{eqFctPhi}
\Phi^n_k(x)= \frac{(-1)^k}{2\pi \I}\oint_{\Gamma_0}\frac{\dx v}{v} \frac{1+2v}{e^{vt}}\frac{(1+v)^{z-1}}{(v(1+v))^{k}}
\end{equation}
where $\Gamma_0$ is an anticlockwise simple loop enclosing only the pole at $0$.
\end{lem}
\begin{proof}
We have
\begin{equation}
\Psi^n_k(x)=(-1)^k F_{-k}(x-y_{n-k},t)= (-1)^k F_{-k}(z-2k,t).
\end{equation}
Then (\ref{eqFn}) leads directly to (\ref{eqFctPsi}). Next we prove that (\ref{eqFctPhi}) satisfy the orthogonality relation (\ref{Sasortho}). Since $\Psi^n_k(x)=0$ for $z(x)<0$, we have
\begin{eqnarray}\label{eq246}
\sum_{z\geq 0}\Psi^n_k(x(z))\Phi^n_j(x(z)) &=&\frac{(-1)^k}{2\pi \I}
\oint_{\Gamma_0}\dx w ^{(w-1) t} ((w-1)w)^k \\ & & \times
\frac{(-1)^j}{2\pi \I}\oint_{\Gamma_0} \frac{\dx v}{v}
\frac{(1+2v)}{e^{vt} (v(v+1))^j} \sum_{z\geq
0}\frac{(v+1)^{z-1}}{w^{z+1}}\nonumber
\end{eqnarray}
provided that the integration domain satisfy $|1+v| < |w|$. The last
sum gives
\begin{equation}
\sum_{z\geq 0}\frac{(v+1)^{z-1}}{w^{z+1}}=\frac{1}{(w-(1+v))(1+v)}.
\end{equation}
Thus (\ref{eq246}) has a simple pole at $w=1+v$, and once the
integral over $w$ is computed, we get
\begin{equation}\label{eq248}
\sum_{z\geq 0}\Psi^n_k(x(z))\Phi^n_j(x(z))= \frac{(-1)^{k+j}}{2\pi \I}
\oint_{\Gamma_0}\dx v \frac{1+2v}{v(1+v)} (v(1+v))^{k-j}.
\end{equation}
The final step is a change of variable. Let $u=v(1+v)$. Then
\begin{equation}
\dx u = (1+2v)\dx v
\end{equation}
and the integral is again around $0$. Thus we get
\begin{equation}
\sum_{z\geq 0}\Psi^n_k(x(z))\Phi^n_j(x(z))= \frac{(-1)^{k+j}}{2\pi \I}
\oint_{\Gamma_0}\dx u \frac{1}{u^{j+1-k}}=\delta_{j,k}.
\end{equation}
\end{proof}

Once the orthogonalization is made, we determine the kernel of
Theorem~\ref{ThmKernel}.
\begin{proofOF}{Theorem~\ref{ThmKernel}}
  We need to derive the formula for the first term (the main part) of
  the kernel. For convenience, we first shift the integrating variable
  of $\Psi^n_k$ to go around $-1$ by setting $u=w-1$. This leads to
\begin{equation}
\Psi^n_k(x)=\frac{(-1)^k}{2\pi \I}\oint_{\Gamma_{-1}}\frac{\dx u}{(1+u)^{z+1}}e^{ut}(u(1+u))^k
\end{equation}
with $z=x+2n-2N$. We start with particles at positions $y_i=2N-2i$, $i=1,\ldots,2N$. The
main term in the kernel writes, with $z_i=x_i+2(n_i-N)$,
\begin{eqnarray}
& &
\sum_{k=0}^{n_2-1}\Psi^{n_1}_{n_1-n_2+k}(x_1)\Phi^{n_2}_k(x_2)=\frac{(-1)^{n_1-n_2}}{(2\pi \I)^2}\oint_{\Gamma_0}\dx v
\frac{(1+2v)(1+v)^{z_2}}{e^{vt}(v(1+v))^{n_2}}\nonumber \\ & & \times
\oint_{\Gamma_{-1}} \dx u \frac{e^{ut} (u(1+u))^{n_1}}{(1+u)^{z_1+1}}
\frac{1}{u(1+u)-v(1+v)}
\end{eqnarray}
provided that the integration paths satisfies (a)
$|u(1+u)|>|v(1+v)|$ and (b) $u=0$ is not inside the contour
$\Gamma_{-1}$.  To obtain this expression we first take the finite
sum over $k$ inside the integrals, and secondly we extend it to
$k=-\infty$. This can be done since the sum is absolutely summable
because of (a) and we do not create new poles inside the
integration contours because of (b). For example, we can set
$\Gamma_1$ by $|1+u|=1/2$ and take $\Gamma_0$ to be a contour with
$|v|$ small enough.

To obtain the kernel for the alternating initial configuration we
focus on the $x_i$'s far enough from the right-most particle so
that the system in the considered region becomes independent of
the fact that we have only a finite number of particles. This is
obtained when $z_i<n_i$, i.e., whenever $u=-1$ is not anymore a
pole. This condition is satisfied for any fixed $x_i$ (i.e.,
around the origin) and any finite time $t$ by taking $N$ large
enough. In fact, it corresponds to taking $n_i-N=\Or(1)$ in $N$.
In this case we are left with one simple pole at $u=-1-v$. Denote
$n_i=N+m_i$, then $z_i=x_i+2m_i$ and the main part of the kernel
becomes, for any $x_i$'s as $N\to\infty$,
\begin{equation}\label{KernelD2}
\sum_{k=0}^{n_2-1}\Psi^{n_1}_{n_1-n_2+k}(x_1)\Phi^{n_2}_k(x_2)=
\frac{-1}{2\pi \I}\oint_{\Gamma_0} \dx v
\frac{(1+v)^{x_2+m_1+m_2}}{(-v)^{x_1+m_1+m_2+1}}e^{-t(1+2v)}.
\end{equation}
Finally, by relabelling the particles we obtain (\ref{eqKernelD2}).
\end{proofOF}

\section{Asymptotic analysis}\label{SectAsympt}
In this section we do the asymptotic analysis for the alternating
initial conditions and prove Theorem~\ref{ThmConvPt}. Just to remind,
the scaling limit we have to consider is
\begin{eqnarray}
x_i&=&-2 u_i t^{2/3}- s_i t^{1/3},\nonumber \\ n_i&=&t/4+ u_i t^{2/3}.
\end{eqnarray}

\begin{proofOF}{of Theorem~\ref{ThmConvPt}}
  The pointwise limit of the first term is quite easy to obtain. Let
  us set $a=(u_2-u_1)t^{2/3}-1$, $b=(s_2-s_1)t^{1/3}+1$, and $\e=b/a$.
  Then we have to compute
\begin{equation}
t^{1/3} \binom{a(2+\e)}{a}.
\end{equation}
We simply use $x!= \sqrt{2\pi x}\exp(x\ln(x)-x)(1+\Or(x^{-1}))$. Since
for $s_1,s_2$ is a bounded set, $\e\to 0$ as $t\to\infty$, we have
that
\begin{equation}
t^{1/3} \binom{a(2+\e)}{a}= t^{1/3} 2^{x_1-x_2}\frac{1}{\sqrt{4\pi
a}}\exp(-b^2/4a)(1+\Or(\e))
\end{equation}
and by replacing back the expressions of $a$ and $b$ we get,
\begin{equation}
\lim_{t\to\infty}t^{1/3}\binom{x_1-x_2-1}{n_2-n_1-1}
2^{x_2-x_1}=\frac{1}{\sqrt{4\pi(u_2-u_1)}}
\exp\left(-\frac{(s_2-s_1)^2}{4(u_2-u_1)}\right).
\end{equation}

Next we analyze the second of the kernel (\ref{eqKernelD2}) multiplied
by $t^{1/3}$. This writes
\begin{equation}
\frac{-t^{1/3}}{2\pi \I} \oint_{\Gamma_0}\dx v \exp\left(t
f_0(v)+t^{2/3} f_1(v)+ t^{1/3} f_2(v)+f_3(v)\right)
\end{equation}
with
\begin{eqnarray}
f_0(v)&=&\frac{1}{2}\ln\left(\frac{1+v}{-v}\right)-1-2v,\nonumber \\
f_1(v)&=& -(u_2-u_1) \ln\left((1+v)(-v)\right),\nonumber \\ f_2(v)&=&
-s_2 \ln(1+v)+s_1\ln(-v),\nonumber \\ f_3(v)&=& -\ln(-v).
\end{eqnarray}
To do a steep descent analysis we first have to find the stationary
points of $f_0(v)$. Simple computations lead to
\begin{equation}
\Dt{f_0(v)}{v} = -\frac{(1+2v)^2}{2v(1+v)},
\end{equation}
which has a double zero at $v=-1/2$. Moreover,
\begin{equation}
\Dtn{f_0(v)}{v}{2}\Big|_{v=-1/2}=0,\quad
\Dtn{f_0(v)}{v}{3}\Big|_{v=-1/2}=16.
\end{equation}

The steep descent path $\Gamma_0$ used for the analysis, shown in
Figure~\ref{FigIntPath}, is given by
$\Gamma_0=\Gamma_0^1\vee\Gamma_0^2\vee\Gamma_0^3$ with
\begin{eqnarray}
\Gamma_0^1&=&\big\{v=-\tfrac12+w e^{-\I\pi/3}, w\in
[0,1/2]\big\},\nonumber \\ \Gamma_0^2&=&\big\{v=-\tfrac12 e^{\I\theta},
\theta \in [\pi/3,5\pi/3]\big\},\\
\Gamma_0^3&=&\big\{v=-\tfrac12+(1/2-w) e^{\I\pi/3}, w\in
[0,1/2]\big\}.\nonumber
\end{eqnarray}
\begin{figure}[t!]
\begin{center}
  \psfrag{a}[r]{$-1/2$} \psfrag{q}[l]{$\pi/3$}
  \psfrag{C}[l]{$\mathbb{C}$} \includegraphics[height=4cm]{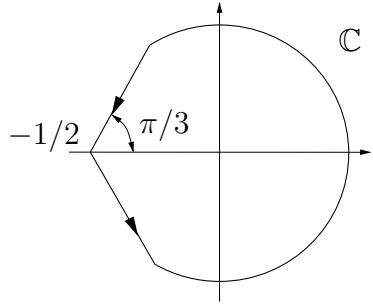}
\caption{The steep descent path $\Gamma_0$ used in the asymptotic analysis.}\label{FigIntPath}
\end{center}
\end{figure}
Let us verify that $\Gamma_0$ is actually a steep descent
path\footnote{For an integral $I=\int_\gamma \dx z e^{t f(z)}$, we
say that $\gamma$ is a steep descent path if (1) $\Re(f(z))$ is
maximum at some $z_0\in\gamma$: $\Re(f(z))< \Re(f(z_0))$ for
$z\in\gamma\setminus\{z_0\}$ and (2) $\Re(f(z))$ is monotone along
$\gamma$ except, if $\gamma$ is closed, at a single point where
$\Re(f)$ reaches its minimum.}. On $\Gamma_0^2$,
\begin{equation}
\Dt{\Re(f_0)(\theta)}{\theta}=-\frac{4\sin\theta
(1-\cos\theta)}{5-4\cos\theta}.
\end{equation}
Therefore the real part of $f_0$ is stationary only at $\theta=0,\pi$
and the maximum is at $\theta=0$, the minimum at $\theta=\pi$. By
symmetry we need to check only on $\Gamma_0^1$. We find
\begin{equation}
\Dt{\Re(f_0)(w)}{w}=-\frac{8w^2(1+2w^2)}{(1+2w+4w^2)(1-2w+4w^2)}
\end{equation}
which is strictly negative except at $w=0$ where is zero. Thus
$\Gamma_0$ is a steep descent path.

Consider now the piece of the path $\Gamma_0^\delta=\{z\in\Gamma^0 |
|z+1/2|\leq \delta\}$. Let us denote $F(v)=\exp(t f_0(v)+t^{2/3}
f_1(v)+ t^{1/3} f_2(v)+f_3(v))$. Then since $\Gamma_0$ is a steep
descent path,
\begin{eqnarray}
\frac{-t^{1/3}}{2\pi \I}\oint_{\Gamma_0}\dx v F(v) &=&
 \frac{-t^{1/3}}{2\pi \I} \int_{\Gamma_0^\delta}\dx v F(v) +
 \frac{-t^{1/3}}{2\pi \I} \int_{\Gamma_0\setminus\Gamma_0^\delta}\dx v
 F(v)\nonumber \\ &=& \frac{-t^{1/3}}{2\pi \I}
 \int_{\Gamma_0^\delta}\dx v F(v) + F(-1/2)\Or(e^{-\mu t})
\end{eqnarray}
for some $\mu>0$ (in our case, $\mu\sim \delta^3$). The precise expression of $F(-1/2)$ is
\begin{equation}
F(-1/2)=2^{2(u_2-u_1)t^{2/3}+(s_2-s_1)t^{1/3}+1}=2^{x_1-x_2+1}.
\end{equation}

For the integral on $\Gamma_0^\delta$ we can apply Taylor development. On
$\Gamma_0^\delta$ we have $v=-1/2+ e^{\pm i\pi/3}w$, $0\leq w \leq
\delta$, and we obtain
\begin{eqnarray}
f_0(v)&=&-\frac83 w^3 + \Or(w^4),\nonumber \\ f_1(v)&=&2
(u_2-u_1)\ln{2}+4 (u_2-u_1) e^{2\pi \I/3} w^2 + \Or(w^3),\nonumber \\
f_2(v)&=& (s_2-s_1) \ln{2} - 2 (s_1+s_2) e^{\I\pi/3} w +
\Or(w^2),\nonumber \\ f_3(v)&=& \ln{2}+\Or(w)
\end{eqnarray}
where the error terms $\Or(\cdots)$ are uniform for $s_1,s_2$ in a
bounded set. Set $\tilde f_i(v)$ to be the expressions $f_i(v)$
\emph{without} the error terms, and similarly $\widetilde F(v)$:
$\widetilde F(v)=\exp\left(t \tilde f_0(v)+t^{2/3} \tilde f_1(v)+
  t^{1/3} \tilde f_2(v)+\tilde f_3(v)\right)$. Then
\begin{equation}
\frac{-t^{1/3}}{2\pi \I}\int_{\Gamma_0^\delta}\dx v F(v) =
\frac{-t^{1/3}}{2\pi \I} \int_{\Gamma_0^\delta}\dx v \widetilde F(v) +
\frac{-t^{1/3}}{2\pi \I}\int_{\Gamma_0^\delta}\dx v (F(v)-\widetilde
F(v)).
\end{equation}
To estimate the second integral, we use the inequality $|e^x-1|\leq
e^{|x|}|x|$. Thus
\begin{eqnarray}\label{eq5.16}
& & \Big|\frac{-t^{1/3}}{2\pi \I} \int_{\Gamma_0^\delta}\dx v
(F(v)-\widetilde F(v)) \Big|\leq \frac{t^{1/3}}{\pi} \int_0^\delta\dx
w|\widetilde F(v(w)=e^{-\I\pi/3}w-1/2)| \nonumber \\ &\times&
e^{\Or(w^4 t+w^3 t^{2/3}+w^2 t^{1/3}+w)}\Or(w^4 t+w^3 t^{2/3}+w^2
t^{1/3}+w)\nonumber \\ &=& \frac{t^{1/3}}{\pi} \int_0^\delta\dx w
|e^{t \tilde f_0(v(w))(1+\chi_1)+t^{2/3}\tilde f_1(v(w)) (1+\chi_2)+
t^{1/3}\tilde f_2(v(w)) (1+\chi_3)}|\nonumber \\ &\times & \Or(w^4
t+w^3 t^{2/3}+w^2 t^{1/3}+w)
\end{eqnarray}
for some $\chi_1,\chi_2,\chi_3$ which can be made as small as
desired by choosing $\delta$ small enough. At the integration
boundary $w=\delta$ the leading term is \mbox{$\exp(-8 \delta^3 t (1+\chi_1)/3)$}. This easily implies that the integral remains
bounded as $t\to\infty$. Now we do the change of variable $z=t^{1/3} w$.
The $t^{1/3}\dx w=\dx z$ and $\Or(w^4t+...+w)=\Or(z^4+...+z)t^{-1/3}$.
The rest of the integral is $e^{-\frac83 z^3(1+\chi_1)+c_1 z^2+c_2 z}$ for some constants $c_1,c_2$, since
the function $\tilde f_i$ do not contains the error terms. The integrand is then $t^{-1/3}$ times $\Or(z^4+...+z)e^{-\frac83 z^3(1+\chi_1)+c_1 z^2+c_2 z}$ and the integral is on $[0,\delta t^{1/3}]$. The $e^{-\frac83 z^3(1+\chi_1)}$ dominates the integral for large $z$. Thus $t^{1/3}\times(\ref{eq5.16})$ remains finite in the $t\to\infty$ limit. Therefore the above estimate of the error term becomes $F(-1/2)\Or(t^{-1/3})$.

The final step is to compute $\frac{-t^{1/3}}{2\pi
\I}\int_{\Gamma_0^\delta}\dx v \widetilde F(v)$. Extending
$\delta$ to $\infty$ we only make an error of order
$F(-1/2)\Or(e^{-\mu t})$ for some $0<\mu\sim \delta^3$ and this
leads to the integration along the path
$\gamma_{\infty}=\{e^{-\I\pi\sgn(w)/3}|w|, w\in \R\}$. Therefore
\begin{eqnarray}
& & \frac{-t^{1/3}}{2\pi \I} \int_{\Gamma_0^\delta}\dx v \widetilde
F(v) = F(-1/2) \frac{-t^{1/3} e^{-\I\pi/3}}{2\pi \I}\int_{0}^\infty\dx w
2 e^{-8 w^3 t/3} \nonumber \\ &\times & e^{4(u_2-u_1) w^2
t^{2/3}e^{-2\pi \I/3}} e^{-2(s_1+s_2)w t^{1/3} e^{-\I\pi/3}} +
F(-1/2)\Or(e^{-\mu t}).
\end{eqnarray}
The change of variable $z=2 t^{1/3} e^{-\I\pi/3} w$ leads then to
\begin{eqnarray}
\frac{-t^{1/3}}{2\pi \I} \int_{\Gamma_0^\delta}\dx v \widetilde F(v)
&=& \frac{ F(-1/2)}{-4\pi \I}\int_{\gamma_\infty}\dx z
e^{z^3/3+(u_2-u_1)z^2-(s_1+s_2)z} \nonumber \\ & &+ F(-1/2)\Or(e^{-\mu
t}).
\end{eqnarray}
Finally we use an Airy function representation
\begin{equation}
\frac{1}{-2\pi \I}\int_{\gamma_\infty}\dx v
e^{v^3/3+av^2+bv}=\Ai(a^2-b)\exp(2a^3/3-ab)
\end{equation}
to obtain the final result
\begin{equation}
\lim_{t\to\infty}\frac{2}{F(-1/2)} \int_{\Gamma_0}\dx v F(v) =
\Ai(s_1+s_2+(u_2-u_1)^2)e^{2(u_2-u_1)^3/3+(s_1+s_2)(u_2-u_1)}.
\end{equation}
\end{proofOF}

\appendix
\section{Compact form for the extended kernel}\label{AppCompactKernel}
In this Appendix we show that the entries of the compact form of
the kernel (\ref{eqKernelCompact}) agree with
(\ref{KernelAsympt}). Let us introduce some notations. Let $Q$ be
the multiplication operator by the position, $D$ be the
differentiation operator, and let $\Delta$ be the Laplacian. On
Schwarz test functions $f\in {\cal S}(\R)$,
\begin{equation}
(Qf)(x)=xf(x),\quad (Df)(x)=\frac{\partial}{\partial x}f(x),\quad (\Delta f)(x)=\frac{\partial^2}{\partial x^2}f(x).
\end{equation}
Moreover, denote by $K_\lambda$ the operator with kernel $K_\lambda(x,y)=\Ai(x+y+\lambda)$ and the Airy operator $H_A=-\Delta+Q$.

We will apply Baker-Campbell-Haussdorf formula. If $[A,[A,B]]=c\Id$ and $[B,[A,B]]=c'\Id$ for some constant $c,c'$, then
\begin{equation}\label{eqBCH}
e^A e^B = e^{A+B+\tfrac12 [A,B]+\tfrac{1}{12}[A,[A,B]]-\tfrac{1}{12}[B,[A,B]]}
\end{equation}
from which follows, for $[A,B]=c\Id$,
\begin{equation}\label{eqBCH2}
e^{A+B}=e^A e^B e^{-\tfrac12 [A,B]}.
\end{equation}
Moreover, we will use the property of Airy functions
\begin{equation}\label{eqAiry}
\Ai''(x+y)=(x+y)\Ai(x+y).
\end{equation}
We collect some useful properties in the following lemma.
\begin{lem}\label{lem3}$ $
\begin{enumerate}
\item[1)] Commutation relations: $[Q,D]=-\Id$,\quad $[Q,\Delta]=-2D$,
\item[2)] $H_A K_\lambda= -K_\lambda (\lambda \Id+Q)$,
\item[3)] $e^{t D} K_\lambda = K_{\lambda+t}$ and $K_\lambda e^{t D} = K_{\lambda-t}$,
\item[4)] $e^{-t \Delta} K_\lambda = e^{-\tfrac23 t^3-\lambda t} e^{-tQ} K_{\lambda+t^2} e^{-tQ}$.
\end{enumerate}
\end{lem}
\begin{proof}

1) By applying to $f\in {\cal S}(\R)$ we get $[Q,D]f(x)=-f(x)$ and $[Q,\Delta]f(x)=-2f'(x)=-2Df(x)$.\\
2) We apply the definition of $H_A$ and use (\ref{eqAiry}) to get
\begin{equation}
(H_A K_\lambda f)(x)=-\int\dx y \Ai(x+y+\lambda)(y+\lambda) f(y)=-(K_\lambda (\lambda \Id+Q)f)(x).
\end{equation}
3) Follows from $(e^{tD}f)(x)=f(x+t)$.\\
4) We use the property 2) and (\ref{eqBCH}) with $A=-t \Delta$ and $B=-t(-\Delta+Q)$, to get
\begin{equation}
e^{-t\Delta}K_\lambda = e^{-t\Delta} e^{-t(-\Delta+Q)} e^{t H_A} K_\lambda
= e^{-\tfrac16 t^3} e^{-tQ+t^2 D} K_\lambda e^{-tQ} e^{-\lambda t}
\end{equation}
Then apply (\ref{eqBCH2}) with $A=-tQ$ and $B=t^2 D$ to get 4).
\end{proof}
What we have to compute explicitly is $e^{-u_1\Delta}K_0 e^{u_2\Delta}$. From 4) of Lemma~\ref{lem3} we have
\begin{equation}\label{eqA7}
e^{-u_1\Delta}K_0 e^{u_2\Delta} = e^{-\tfrac23 u_1^3} e^{-u_1Q} K_{u_1^2} e^{-u_1Q} e^{u_2\Delta}.
\end{equation}
The last part can be rewritten as
\begin{equation}
e^{-u_1Q} e^{u_2\Delta} = e^{(u_2\Delta -u_1 Q-\tfrac16 u_2 u_1^2-u_1 u_2 D) + (2u_1u_2 D)}
\end{equation}
by (\ref{eqBCH}) with $A=-u_1Q$ and $B=u_2\Delta$. Then using (\ref{eqBCH2}) with \mbox{$A=u_2\Delta -u_1 Q-\tfrac16 u_2 u_1^2-u_1u_2 D$} and $B=2u_1u_2D$ we obtain
\begin{equation}
e^{-u_1Q} e^{u_2\Delta} = e^{2u_1u_2 D} e^{u_2\Delta} e^{-u_1Q} e^{u_2u_1^2}.
\end{equation}
Plugging this back into (\ref{eqA7}) we have
\begin{equation}
e^{-u_1\Delta}K_0 e^{u_2\Delta} =e^{-\tfrac23 u_1^3+u_2u_1^2} e^{-u_1Q} K_{u_1^2} e^{2u_2u_1 D} e^{u_2\Delta} e^{-u_1Q}.
\end{equation}
Then we apply 3) of Lemma~\ref{lem3}, namely $K_{u_1^2} e^{2u_2u_1 D}=K_{u_1^2-2u_2u_1}$, and we exchange the order of $K_\cdot$ and $e^{u_2\Delta}$ because are both symmetric and apply 4) of Lemma~\ref{lem3}. This results into
\begin{equation}
e^{-u_1\Delta}K_0 e^{u_2\Delta} = e^{-\tfrac23(u_1-u_2)^3} e^{-(u_1-u_2) Q} K_{(u_1-u_2)^2} e^{-(u_1-u_2)Q}.
\end{equation}
Explicitly
\begin{equation}
(e^{-u_1\Delta}K_0 e^{u_2\Delta})(s_1,s_2) = e^{\tfrac23(u_2-u_1)^3} e^{(u_2-u_1)(s_1+s_2)} \Ai(s_1+s_2+(u_1-u_2)^2).
\end{equation}
Thus we showed how the second term in (\ref{eqKernelCompact})
leads to the corresponding one in (\ref{eqKernelExpanded}). It
remains the first one, $(e^{(u_2-u_1)\Delta})(s_1,s_2)$, for
$u_2>u_1$. This is just the one-dimensional heat kernel, for which
it is well known that (see e.g.~\cite{Oks98})
\begin{equation}
(e^{(u_2-u_1)\Delta})(s_1,s_2)=\frac{1}{\sqrt{4\pi (s_2-s_1)}}\exp\left(-\frac{(u_2-u_1)^2}{4(s_2-s_1)}\right).
\end{equation}

\section{Charlier polynomials}\label{AppCharlier}
In this Appendix we explain a constructive method to do the
orthogonalization.  Let $C_n(x,t)$ be the Charlier polynomial of
degree $n$. They are orthogonal polynomials with respect to the
weight on $\{0,1,\ldots\}$ given by
\begin{equation}
w_t(z)=e^{-t} t^z/z!
\end{equation}
which are traditionally normalized via
\begin{equation}
\sum_{z\geq 0} C_n(z,t) C_m(z,t) w_t(z)=\frac{n!}{t^n}\delta_{n,m}
\end{equation}
or, equivalently, $C_n(z,t)=(-1/t)^n z^n+\cdots$.  They can be
expressed in terms of hypergeometric functions
\begin{equation}
C_n(x,t)=\phantom{}_2F_0(-n,-x;\, ;-1/t)
\end{equation}
and satisfy the recurrence relation
\begin{equation}\label{eqApp3}
\frac{x}{t} C_n(x-1,t)=C_n(x,t)-C_{n+1}(x,t).
\end{equation}
From the generating function of the Charlier polynomials
\begin{equation}
\sum_{n\geq 0}\frac{C_n(x,t)}{n!}v^n=e^v(1-v/t)^x
\end{equation}
one gets the integral representation
\begin{equation}
\frac{1}{n!} C_n(z,t)=\frac{1}{2\pi \I}\oint_{\Gamma_0}\frac{dv}{v}
\frac{e^v(1-v/t)^z}{v^n}.
\end{equation}
For a good reference on orthogonal polynomials, see~\cite{KS96}.

It is not too difficult to see that the functions $\Psi_k^{N}$
defined in Lemma~\ref{lemOrtho} can be expressed in terms of the
Charlier orthogonal polynomials as
\begin{equation}
\Psi_k^{N}(z)=\frac{e^{-t}t^{z-k}}{(z-k)!}C_k(z-k,t).
\end{equation}
Using the recurrence relation (\ref{eqApp3}) repeatedly we obtain
\begin{equation}
\Psi_k^{N}(z)=w_t(z) \sum_{l=0}^{2 k} S_{k,l} C_l(z,t),
\end{equation}
where the entries of the matrix $S$ are
\begin{equation}
S_{k,l}=(-1)^{l-k}\binom{k}{l-k}.
\end{equation}
Notice that $S$ is not a square matrix.  From this it follows that the
polynomials $\Phi_k^{N}$ which satisfy
\begin{equation}
\sum_{z\geq 0} \Phi_k^{N}(z) \Psi_j^{N}(z)=\delta_{k,j},
\end{equation}
are given by
\begin{equation}\label{eqApp1}
\Phi_k^{N}(z)=\sum_{l=0}^{N-1} C_l(z,t) \frac{t^l}{l!} \tilde
S^{-1}_{l,k}.
\end{equation}
where by $\tilde S^{-1}_{i,j}$ we mean the $(i,j)$-entry of the
inverse of the square matrix $\tilde S=[S_{i,j}]_{0\leq i,j \leq
N-1}$ obtained by restricting $S$ to the first $N$ indices. The
first main difficulty is to obtain the inverse of $\tilde S$.
After some work we could determine it, namely
\begin{equation}\label{eqApp2}
\tilde S^{-1}_{i,j}=\binom{2j-i}{j-i}\frac{i}{2j-i}
\end{equation}
with the identification $\tilde S^{-1}_{0,0}=1$ and the convention
that the RHS of (\ref{eqApp2}) is zero when $i>j$.

At this point we substitute (\ref{eqApp2}) into (\ref{eqApp1}),
perform the summation, and finally change the variable $v=-wt$.
The final result is the biorthogonal functions $\Phi_k^{N}$
reported in (\ref{eqFctPhi}).


\end{document}